\documentstyle[12pt,tighten,aps,epsfig]{revtex}
\begin{document}
\draft
\title{
Synchronisation and MSW sharpening of neutrinos
propagating in a flavour blind medium.}
\author{Nicole F. Bell$^1$, R. F. Sawyer$^2$ and Raymond R. Volkas$^1$ }
\address{$^1$ School of Physics, Research Centre for High Energy Physics\\
The University of Melbourne, Victoria 3010 Australia\\
$^2$ Department of Physics, University of California at Santa Barbara \\
Santa Barbara, California 93106\\
(n.bell@physics.unimelb.edu.au, sawyer@vulcan.physics.ucsb.edu, 
r.volkas@physics.unimelb.edu.au)}
\maketitle

\begin{abstract}

We consider neutrino oscillations in a medium in which scattering processes 
are blind to the neutrino flavour.  We present an analytical derivation of the 
synchronised behaviour obtained in the limit where the average scattering rate is
much larger than the oscillation frequency.
We also examine MSW transitions in these circumstances, and show that a 
sharpening of the transition can result.

\end{abstract}

\section{Introduction}

We examine a system consisting of an oscillating neutrino propagating
through a dense medium. This falls into the very broad class of problems
in which a quantum system cannot be considered closed or isolated, due to
interaction of the system with an environment or heat bath. In particular
we consider two-flavour neutrino oscillations in a medium such that the
system-bath interactions do not distinguish the flavour of the neutrino.

An example is the density matrix of a single neutrino
undergoing $\nu_{\mu}-\nu_{\tau}$ oscillations embedded in an
electron-positron plasma. 
In some sense this example is similar to an early universe environment, 
though lacking some features of a realistic physical scenario,
such as neutrino pair processes and the inclusion of other neutrino flavours.
Another example in which flavour blind interactions are relevant
is the case of $\nu_{\mu}-\nu_{\tau}$ oscillations of neutrinos coming
from the core of a supernova.  In this case, although the indices of
refraction for $\nu_{\mu}$ and $\nu_{\tau}$ will be different due to loop
graph effects, the collision rates for the neutrinos scattering from the
medium of nucleons will be the same for the two flavours, to a good
approximation. 
However, it is the coherence properties of the system that
are most intriguing, and which are worth studying independently of the
various circumstances in which they may arise.

An analogue to the neutrino system is that of a double well coupled to a thermal
bath, in which the neutrino flavour oscillations are
replaced by oscillations from the left to the right side of the well.  
Our work provides an analytic treatment of the distinctive behaviour that
we noted previously in a numerical treatment of double well oscillations 
\cite{doublewell}.
Since double well systems are of general interest in condensed matter 
physics \cite{leggett,nature}, the type of behaviour we examine here may be 
expected to arise in various situations unrelated to neutrino physics.

As the medium cannot distinguish the flavour of the neutrino, the only
effect of collisions with particles in the medium is to change the value
of the neutrino's energy (and to entangle the neutrino with the environment). 
This is in contrast with active-sterile
oscillations in which collisions may in some sense be thought of as
measuring the flavour of the neutrino. It is the consideration of the
neutrino energy variable, and the rapid energy changing collisional
processes that is crucial, and when the average collision rate is much 
larger than average oscillation rate we find
that a collective oscillation results \cite{doublewell}.  That is, for 
a neutrino in a thermally distributed mixture of momentum states, the
oscillations are synchronised despite being momentum dependent in the
absence of collisions.

We stress that the novel synchronisation effect arises as a result of energy 
changing processes.  In particular, the result is distinct from the 
synchronisation effect of ref.\cite{synchro} which results from non-linear 
feedback, as we consider here only linear dynamics.

\subsection{The Quantum Kinetic Equations}

The evolution equations for the neutrino density matrix have
been formulated in ref.\cite{active} for active-active oscillations, and 
ref.\cite{sterile} for active-sterile oscillations.  
These Quantum Kinetic Equations (QKEs) or Bloch equations,  
take into account both coherent oscillation behaviour and potentially 
decoherent effects of interaction of the neutrino system with an environment.
The QKEs are essentially Boltzmann equations, generalised to allow for quantal 
coherence between particle species.

We choose to parametrise the neutrino density matrix as
\begin{equation}
\rho(p)  =  \frac{1}{2}[P_0(p) + {\bf P}(p) \cdot {\bf \sigma}],
\end{equation}
with
\begin{equation}
{\bf P}(p) \equiv P_x(p) \hat{x}
+ P_y(p) \hat{y} + P_z(p) \hat{z}.
\end{equation}

The QKEs for a single neutrino undergoing elastic scattering in a 
medium which does not distinguish flavour are \cite{active}
\begin{eqnarray}
\label{xyz}
\dot{\vec{P}}(k) &=& 
\int d^3k' \left[ \Gamma(k',k) \vec{P}(k') - \Gamma(k,k') \vec{P}(k) \right]
+ {\bf V}(k) \times {\bf P}(k), 
\nonumber \\
\dot{P}_0(k) &=& \int d^3k' \Bigl[ \Gamma(k',k) P_0(k') - \Gamma(k,k') P_0(k) \Bigr], 
\end{eqnarray}
where $\Gamma(k',k)$ determines the rate at which neutrinos are scattered from 
momentum state $k'$ to state $k$.
Thermal equilibrium of the medium enforces
\begin{equation}
\label{thermal}
\Gamma(k',k)= e^{(k'-k)/T} \Gamma(k,k').
\end{equation}
The cross product term describes precession of the neutrino flavour, 
where ${\bf V}(k)$ is given by
\begin{equation}
{\bf V}(k)= \beta(k) \hat{x} + \lambda(k) \hat{z},
\end{equation}
and the oscillation parameters are defined as 
\begin{eqnarray}
\beta(k) &=& \frac{\delta m^2}{2k} \sin 2 \theta_0, \nonumber \\
\lambda(k) &=& -\frac{\delta m^2}{2k} \cos 2 \theta_0 + V_{\rm eff},
\end{eqnarray}
with $\theta_0$ being the vacuum mixing angle, $\delta m^2$ the mass-squared 
difference, and $V_{\rm eff}$ the effective matter potential resulting from 
coherent forward scattering.

\subsection{Qualitative features for different coupling strength.}

By way of example, we consider the density matrix of a single 
$\nu_{\mu}-\nu_{\tau}$
neutrino propagating in an electron-positron plasma. For simplicity, we
shall ignore the interaction of the neutrino with other $\nu_{\mu,\tau}$ or
$\overline{\nu}_{\mu,\tau}$ that may be produced in the plasma, and consider 
only the single neutrino and its interaction with electrons and positrons.
That is, we consider in isolation a particular subset of the various 
processes that may affect the evolution of the neutrino density matrix.

For our considerations it should suffice to take the following form for 
the scattering rate,
\begin{equation}
\Gamma(k,k') \simeq  \frac{0.01}{4 \pi} G_F^2 k k' e^{-k'/T},
\end{equation}
as this embodies the correct threshold behaviour, the reciprocity relation 
(\ref{thermal}) and leads to the correct total transition rate.
We have solved the QKEs (\ref{xyz}) numerically, and the results are
presented in the graphs below, for various temperatures. For small
temperatures, where the average time between collisions is much longer
than the oscillation period, we have just the superposition of (almost)
undamped precession rates at different frequencies. For moderate
collisions rates we rapidly lose coherence in the neutrino system. This
occurs simply because the oscillation modes are momentum dependent.  
Collisional processes transfer quanta across the momentum states which
quickly get out of phase, and hence result in a damping of the amplitude
of the oscillations. The most fascinating effect, however, is obtained for
larger values of $\Gamma$ where the average time between collisions is much 
smaller than the oscillation period.  In this limit we see a synchronisation 
of the oscillations of all the momentum states. Here, the precession proceeds at
at a rate determined by the thermal average of ${\bf V}(k)$, 
and damping is
eliminated as the rapid collisions force the momentum modes to stay in
phase.

\begin {figure}[h]
    \begin{center}
        \epsfxsize 3in
        \begin{tabular}{rc}
            \vbox{\hbox{
$\displaystyle{ \, {  } } $
               \hskip -0.1in \null} %\vskip 1.9in
} &
            \epsfbox{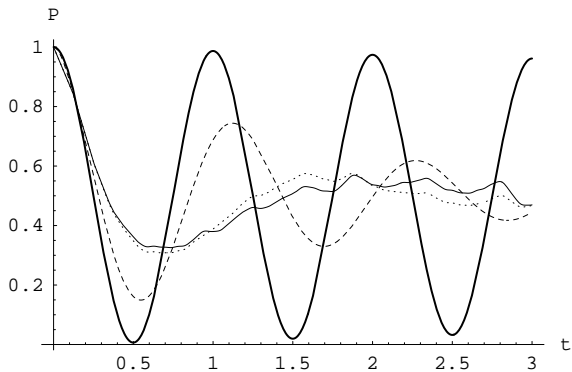} \\
            &
            \hbox{ } \\
        \end{tabular}
    \end{center}
\label{sync}
%\vskip 1in   
\caption
	{The probability that the neutrino is a $\nu_{\mu}$, where 
the initial condition is a $\nu_{\mu}$ with a thermal distribution 
of energies.
The solid, dotted, dashed and heavy solid curves correspond to 
T=5,10,20 and 50MeV respectively, and we 
have taken $\delta m^2=0.001$eV$^2$, $\sin 2 \theta_0=1$.  In each case, the time 
is scaled by the inverse of the thermally averaged oscillation frequency.}
\end{figure}

We define the entropy in the neutrino system as 
\begin{equation}
s=- \int d^3k {\rm Tr}\left[ \rho(k) \ln \rho(k) \right],
\end{equation}
where the trace is over the neutrino flavour index.  
The entropy is plotted in fig.2 for various collision rates.

\begin{figure}[h]
    \begin{center}
        \epsfxsize 3in
        \begin{tabular}{rc}
            \vbox{\hbox{
$\displaystyle{ \, {  } } $
               \hskip -0.1in \null} %\vskip 1.9in
} &
            \epsfbox{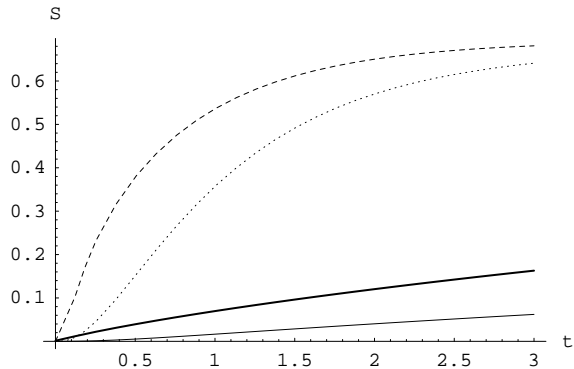} \\
            &
            \hbox{ } \\
        \end{tabular}
    \end{center}
\label{entropy}
%\vskip 1in   
\caption
{Entropy plotted for the same parameters as fig.1, where S=s(t)-s(t=0).}
\end{figure}

Let us compare this behaviour with the familiar Quantum Zeno behaviour in
active-sterile neutrino systems. In both cases the rate of entropy growth
increases as we increase the interaction rate between the neutrino system
and the environment, and then decreases again if we make the interaction
rate much larger than the oscillation rate.  In the active-sterile case
however, this occurs because the ``measurement-like'' interactions freeze
the neutrino in its initial state.  In contrast, in our case the flavour
precession proceeds as though it were a single mode of an isolated
system.

\section{Analytic solution for large collision rate}

We now show analytically that a synchronised solution follows from the 
Eqs. (\ref{xyz}) in the large $\Gamma$ limit.
We begin be discretizing the equations, by allowing the momentum to take 
the values $k_1, k_2,....k_N$.  Expressing the resulting $3N$ equations in 
matrix form we have
\begin{equation}
\dot{\cal{P}} = {\cal M P},
\end{equation}
where ${\cal P}$ is the vector 
\begin{equation}
{\cal P} = (P_{x1},..., P_{xN}, P_{y1},..., P_{yN}, P_{z1},..., P_{zN})^T.
\end{equation}
The matrix $\cal{M}$, has the following block form
\begin{eqnarray}
\cal{M} =  \left( \begin{array}{ccc}
\cal{G} & - \Lambda & 0 \\
\Lambda & \cal{G} & - \cal{B}  \\
0 & \cal{B} & \cal{G}
\end{array} \right),
\end{eqnarray}
where
\begin{eqnarray}
{\cal G}_{ij} &=& \Gamma_{ji} k_j^2 - \delta_{ij} \sum_{k} \Gamma_{ik} k_k^2, \nonumber \\
\Lambda &=& \rm{diag}(\lambda_1,.....\lambda_N), \nonumber \\
{\cal B} &=& \rm{diag}(\beta_1,.....\beta_N).
\end{eqnarray}

To solve for ${\cal P}(t)$, we introduce a matrix $\cal{U}$ which diagonalises
${\cal M}$ such that
\begin{equation}
\cal{P=UQ},
\end{equation} 
and 
\begin{equation}
\label{dQ}
\dot{\cal{Q}}= {\cal M}_{\rm diag} {\cal Q} - {\cal U}^{-1} \dot{\cal U} {\cal Q},
\end{equation}
where 
\begin{equation}
{\cal M}_{\rm diag} \equiv  {\cal U}^{-1} {\cal M U}.
\end{equation}
We shall consider time independent $\lambda$ and $\beta$, so that 
$\dot{\cal U}=0$, and the solution to eq.(\ref{dQ}) is 
\begin{equation}
{\cal P}_i(t) = \sum_j {\cal U}_{ij} {\cal Q}_j(t) = \sum_j
\exp\left[({\cal M}_{\rm diag})_{jj} t \right] {\cal U}_{ij} {\cal Q}_j(0).
\end{equation}  
The procedure may also be adapted for time dependent 
mixing where $\lambda$ and $\beta$ change sufficiently slowly that an 
adiabatic approximation is valid.\footnote{An adiabatic approximation was applied 
to the case of partially coherent oscillations in Ref.\cite{adiabatic}.}

Since wish to find the solution in the large $\Gamma$ limit, we first drop the 
$\lambda$ and $\beta$ terms.  Then we see that $\cal{M}$ has three zero 
eigenvalues, with eigenvectors that are linear  combinations of 
\begin{equation}
{\bf y_1}=\frac{1}{\sqrt{N}}\left(e^{-k/T}, 0, 0\right)^T, \;\; 
{\bf y_2}=\frac{1}{\sqrt{N}}\left(0,e^{-k/T}, 0\right)^T \;\; 
{\rm and} \;\; {\bf y_3}=\frac{1}{\sqrt{N}}\left(0, 0, e^{-k/T}\right)^T,
\end{equation}
where $e^{-k/T}$ denotes $(e^{-k_1/T}, e^{-k_2/T},...,e^{-k_N/T})$.
The corresponding eigenvectors of ${\cal M}^T$ are
\begin{equation}
{\bf x_1}=\frac{1}{\sqrt{N'}}\left(k^2, 0, 0 \right)^T, \;\; 
{\bf x_2}=\frac{1}{\sqrt{N'}}\left(0,k^2, 0 \right)^T \;\; 
{\rm and} \;\; {\bf x_3}=\frac{1}{\sqrt{N'}}\left(0, 0, k^2 \right)^T,
\end{equation}
with  ${\cal U U}^{-1}=1$ requiring $\sqrt{NN'}= \sum_{k} k^2 e^{-k/T}$.
\footnote{Note that since $\cal{M}$ is neither hermitian nor symmetric, 
U will neither be unitary nor orthogonal.}
All the other eigenvalues are negative and proportional to $\Gamma$.  
To observe this
let $\Omega$ be an eigenvalue of ${\cal G}^T$ with eigenvector ${\bf X}$.  The eigenvalue 
may be expressed as
\begin{equation}
\Omega = \sum_{k} \Gamma_{jk} k_k^2 \left( \frac{X_k k_j^2}{X_j k_k^2} -1  \right), \;\;\;\; 
\forall j  : X_j \neq 0.
\end{equation}
Since we may choose $j$ such that $ | X_j /k_j^2 | \geq  | X_k / k_k^2
|,\; \forall k $, 
it follows that ${\rm Re } \Omega \leq 0$.
Thus the eigenvectors corresponding to the non-zero eigenvalues will make a negligible 
contribution to the solution in the  large $\Gamma$ limit (for $t > 1/ \Gamma$). 
Then the solution is simply
\begin{equation}
{\bf P}_i (t) \simeq e^{-k_i/T} \times \rm{const} \equiv \rm{thermal} \; \rm{solution}.
\end{equation}

Treating $\lambda$ and $\beta$ as perturbation parameters, we are led to consider 
the $3 \times 3$ matrix $V_{ij} = {\bf x}_i^T {\cal V} {\bf y}_j$, where ${\cal V}$
is the matrix obtained by setting ${\cal G}=0$ in  ${\cal M}$.
Using standard degenerate perturbation theory, to first order the three 
eigenvalues become 
\begin{equation}
\pm i w, \;\; 0
\end{equation}
where $\omega$, the synchronised oscillation frequency, 
is\footnote{If one assumes the existence of a synchronised solution,
the oscillation frequency must be as given in eq.(\ref{omega}), 
from the self-consistency of eq.(\ref{xyz}).}

\begin{equation}
\label{omega}
\omega=\sqrt{ \langle \lambda \rangle^2 +  \langle \beta \rangle^2 }
\end{equation}
and $\langle \lambda \rangle$ and $ \langle \beta \rangle $ are the thermally 
averaged values
\begin{equation}
\langle \lambda \rangle  \equiv \frac{\sum_{k}k^2 e^{-k/T}
\lambda_k}{\sum_{k'}k'^2 e^{-k'/T}},  \;\;\;
\langle \beta \rangle  \equiv \frac{\sum_{k}k^2 e^{-k/T}
\beta_k}{\sum_{k'}k'^2 e^{-k'/T}}.
\end{equation}
Note that in the absence of a matter potential, the synchronised
precession frequency is just $\omega=\langle \delta m^2/2k \rangle$. 
For time independent mixing the solution becomes,
\begin{eqnarray}
P_x(k,t) &=& n e^{-k/T}c_{2\Theta}s_{2\Theta}[A'-A\cos(\omega t -\alpha)]
\nonumber \\
P_y(k,t) &=& -n e^{-k/T}A s_{2\Theta} \sin(\omega t-\alpha)
\nonumber \\
P_z(k,t) &=& n e^{-k/T}[A'c^2_{2\Theta}+As^2_{2\Theta}\cos(\omega t -
\alpha)], 
\end{eqnarray}
where
\begin{eqnarray}
c_{2\Theta}&=& \langle \lambda \rangle /w ,\nonumber\\
s_{2\Theta}&=& \langle \beta \rangle /w ,\nonumber\\
n &=& \frac{1}{\sum_{k}k^2 e^{-k/T}} \sum_{k'}k'^2 P_z(k',t=0).
\end{eqnarray}
That is, we have undamped oscillations with frequency $\omega$ at an effective mixing 
angle of $\Theta$.
If the initial condition is such that the neutrino is in the pure flavour state described 
by $(1+\sigma_3)/2$ then $\alpha=0$ and $A'=A=1$.  
Allowing for general initial conditions, such as an initial  
distribution of phases, we have
\begin{eqnarray}
\tan \alpha &=& 
\frac{\sum_{k} k^2
P_y(k,0)}{\sum_{k'}k'^2[s_{2\Theta}P_z(k',0)-c_{2\Theta}P_x(k',0)]}
\nonumber\\
A' &=&  1+ \frac{s_{2\Theta} \sum_{k}k^2P_x(k,0)}{c_{2\Theta}
\sum_{k'}k'^2P_z(k',0)}\nonumber\\
A &=& \left[ \left( \frac{ \sum_{k}k^2P_y(k,0)}{s_{2\Theta}
\sum_{k'}k'^2P_z(k',0)} \right)^2 
+ \left(1-\frac{c_{2\Theta} \sum_{k}k^2P_x(k,0)}{s_{2\Theta}
\sum_{k'}k'^2P_z(k',0)} \right)^2  \right]^{\frac{1}{2}}
\end{eqnarray}

\section{MSW Sharpening}

We now consider the case where there is an MSW resonance, but the
collision rates are still flavour blind. For example, imagine that our 
plasma has a nucleon component, or a lepton excess.
The effective potential arises at 1 loop level, and is
given by \cite{marciano}
\begin{equation}
V= \frac{3 G_F^2 m_{\tau}^2}{2 \pi^2} \left[ (N_p+N_n) \ln
\left( \frac{m_W^2}{m_{\tau}^2} \right) -N_p - \frac{2}{3}N_n \right].
\end{equation}
We need only a very small nucleon component for this potential to be
significant.  For example, assuming a mass squared difference of 
$\delta m^2 \sim 10^{-3} {\rm eV}^2$, a nucleon density of 
$N_n \sim 10^{-3} N_e$ would give rise to an MSW resonance at a
temperature of roughly 10 MeV.  If the nucleon density were smaller, 
the resonance would occur at higher temperature. 

To illustrate the type of effect that could arise, we assume, by way of
example, an exponentially decreasing effective matter potential. We take
for our initial condition a neutrino in a definite flavour state with a
thermal distribution of momentum values, and plot below the probability
that the neutrino remains in that initial flavour state. In the absence
of neutrino-environment interaction, each momentum state goes through the
resonance at a different time resulting in a broad transition region,
while for moderate interaction rate the oscillations get dephased and the
neutrino ends up in an equal mixture of the two flavour states.  For
large interaction rate, we see the effect of synchronised behaviour - all
momentum states go through the resonance at the same time, resulting in a
much sharper transition.

\begin {figure}[h]
    \begin{center}
        \epsfxsize 3in
        \begin{tabular}{rc}
            \vbox{\hbox{
$\displaystyle{ \, {  } } $
               \hskip -0.1in \null} %\vskip 1.9in
} &
            \epsfbox{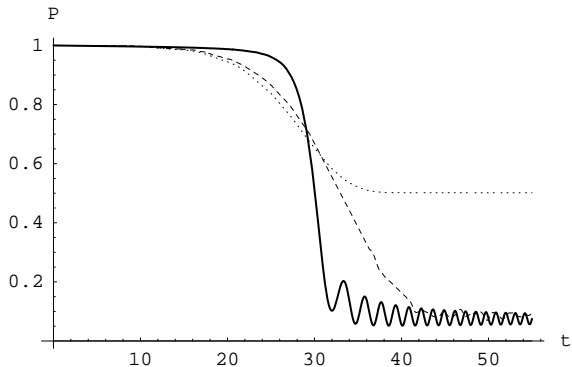} \\
            &
            \hbox{ } \\
        \end{tabular}
    \end{center}
\label{fig2}
%\vskip 1in   
\caption
	{ ``MSW sharpening''.  
The probability that the neutrino remains a $\nu_{\mu}$, given an initial 
thermal distribution of $\nu_{\mu}$.
The dashed, dotted and solid curves are for T=1, 10 and 50 MeV 
respectively, with $\delta m^2=0.001$eV $^2$ and $\sin 2\theta_0=0.2$.
The time is in units of $1/[\omega(V=0)]$. }
\end{figure}

The sharpened MSW resonance when occurs $\langle \lambda \rangle = 0$, 
where the matter potential has the value
\begin{equation}
V_{\rm eff}({\rm res}) 
=\frac{\sum_k k^2e^{-k/T} c_{2\theta_0} \delta m^2/(2k) }{\sum_k k^2 e^{-k/T}}.
\end{equation}

\section{Conclusion}

When rapid scattering processes are flavour blind, a neutrino in a mixture 
of momentum states undergoes synchronised flavour oscillations.
This collective, dissipationless behaviour has been demonstrated numerically
and derived analytically.
Under these circumstances MSW transitions become sharper as all momentum 
states go through the resonance together.

\acknowledgments{
The work of NFB is supported by the Commonwealth of Australia,
RFS by NSF grant PHY-9900544 and RRV by the Australian Research Council.}

\end{document}